\documentclass[%
reprint,
superscriptaddress,
showpacs,
amsmath,amssymb,
aps,
prl,
floatfix,
]{revtex4-1}

\usepackage{hyperref}
\usepackage{graphicx}
\DeclareMathOperator{\sign}{sign}

\DeclareMathOperator{\modd}{mod}
\newcommand{\eps}{\varepsilon}
\newcommand{\dd}{\mathrm{d}}
\newcommand{\varth}{\vartheta}

\errorstopmode

\begin{document}

\preprint{APS/123-QED}

\title{Spherical wedge billiard: from chaos to fractals and Talbot carpets}

\author{Tom\'{a}\v{s} Tyc}%
\email{tomtyc@physics.muni.cz}
\affiliation{Department of Theoretical Physics and Astrophysics, Faculty of Science, Masaryk University, Kotl\'{a}\v{r}sk\'{a} 2, 61137 Brno, Czechia}%

\author{Darek Cidlinsk\'y}%
\affiliation{Department of Theoretical Physics and Astrophysics, Faculty of Science, Masaryk University, Kotl\'{a}\v{r}sk\'{a} 2, 61137 Brno, Czechia}%

\begin{abstract}

We introduce the spherical wedge billiard, a dynamical system consisting of a particle moving along a geodesic on a closed non-Euclidean surface of a spherical wedge. We derive the analytic form of the corresponding Poincar\'{e} map and find very complex dynamics, ranging from completely chaotic to very regular, exhibiting fractal features. Further, we show that upon changing the billiard parameter, the fixed points of the Poincar\'{e} map merge in complex ways, which has origin in the spherical aberration of the billiard mapping. We also analyze in detail the regular regime when phase space diagram is closely related to Talbot carpets. 

\end{abstract}

\maketitle

In the standard billiard problem~\cite{Katok,Gutkin2012}, a particle or light ray is confined to a certain region of a plane where it moves freely with specular reflection at its boundary. The particle dynamics can be regular, completely chaotic or a combination of the two~\cite{Chernov,Andreasen2009}, depending on the shape of the boundary. Billiards play an important role in nonlinear physics and statistical mechanics, are related to many areas of physics~\cite{Douglas,Ponomarenko,friedman2001,gao2015,bittner2012,Bunimovich2005} and have been subject to extensive theoretical and experimental investigation; they are invaluable in investigating relations between the classical dynamics and properties of the corresponding quantum system~\cite{ishikawa1985,cheon1989,Bogomolny2004}.

Consider now a related situation, when the particle or light ray is confined to a closed non-Euclidean surface where it moves freely. As there is no boundary, the particle is not subject to reflection and it continues moving forward along a geodesic, which is the most straight line on the surface. We will call such a system a {\em reflectionless geodesic billiard} (RGB). In contrast to the standard billiard problem, the motion in the RGB is governed by the curvature of the surface rather than by the shape of the boundary. This is a very similar situation as in geodesic lenses that are attracting increasing attention from both the theoretical and experimental point of view~\cite{schultheiss2010,mitchell2014,chen2015,Quevedo-Teruel2022}; there too, it is the curvature of the non-Euclidean surface that yields the desired functionality, e.g. focusing of light, rather than a refractive index profile as in usual lenses. 
The simplest example of RGB is a sphere where all trajectories are periodic and form great circles. Less trivial examples include surfaces of regular polyhedra recently analyzed in~\cite{garcia2020} or a deformed torus~\cite{wang-noneuclidean2020}.

\begin{figure}[thb]
\includegraphics[height=34mm]{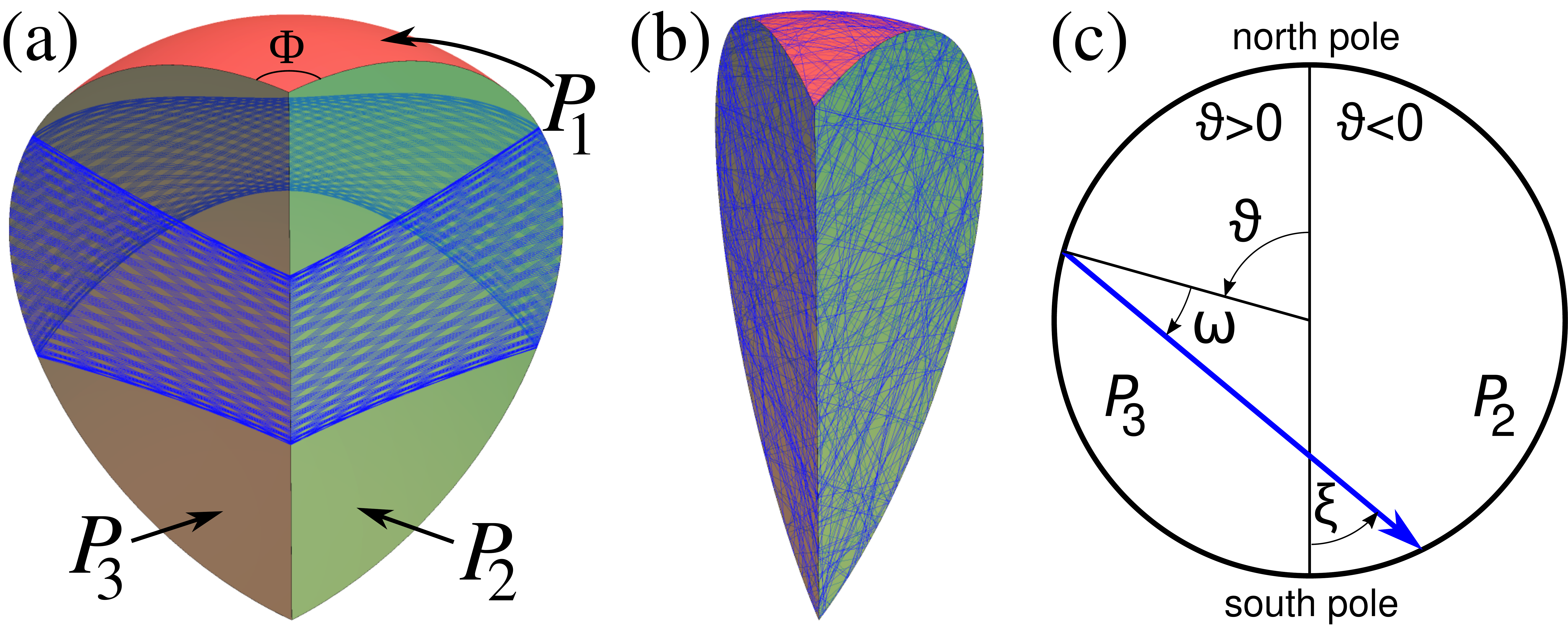}
\caption{The spherical wedge billiard with a sample geodesic trajectory for (a) $\Phi=0.6\,\pi$ and (b) $\Phi=0.25\,\pi$. (c) The angles $\varth$ and $\omega$ labeling the straight particle trajectory segment (blue) on the disk. Alternatively, the segment will be specified with the pair of angles $(\xi,\omega)$.}
\label{fig:manifold}
\end{figure}

Here we present a RGB with the shape of a spherical wedge, see Fig.~\ref{fig:manifold} (a---b); we call it {\em spherical wedge billiard} (SWB). Despite the simplicity of its geometry, the dynamics of this system exhibit extremely complex behavior, ranging from almost completely chaotic to highly regular behavior with fractal features, depending on the value of the billiard parameter $\Phi$. Remarkably, for certain  parameter values the phase space diagram of SWB reflects the structure of the set of rational numbers, closely resembling Talbot carpets known from diffraction theory~\cite{Berry2001}. A great advantage of this system is that the trajectories can be described analytically by formulas containing just trigonometric functions, which allows to analyze many features of the billiard analytically or semi-analytically. 

The spherical wedge billiard consists of three parts (see Fig.~\ref{fig:manifold}(a)): the spherical lune $P_1$ with dihedral angle $\Phi$, and two half-disks $P_2$ and $P_3$ meeting along the polar axis of the spherical lune. We can imagine that we ``unfold'' $P_2$ and $P_3$ into a single disk; the particle will then alternate between moving on the disk and on the spherical lune. Parts of the trajectory on the disk are straight line segments while on the spherical lune they are parts of great circles. When the particle crosses the border between the disk and lune or back, the angle between the trajectory and the border is preserved because the trajectory is a geodesic. The angle $\Phi$ is the sole parameter of the billiard. 

To describe the straight ray segments on the disk, we use a pair of angles $(\varth,\omega)$, see Fig.~\ref{fig:manifold}(c)). Here $\varth$ runs from $-\pi$ to $\pi$ and measures the polar angle on the disk of the point where the particle enters the disk. The $\omega$ runs from $-\pi/2$ to $\pi/2$ and it is the angle between the trajectory and radial line to the entry point; negative  and positive $\omega$ corresponds to the particle running clockwise and counter-clockwise, respectively, around the center.

\begin{figure}[htb]
\includegraphics[width=80mm]{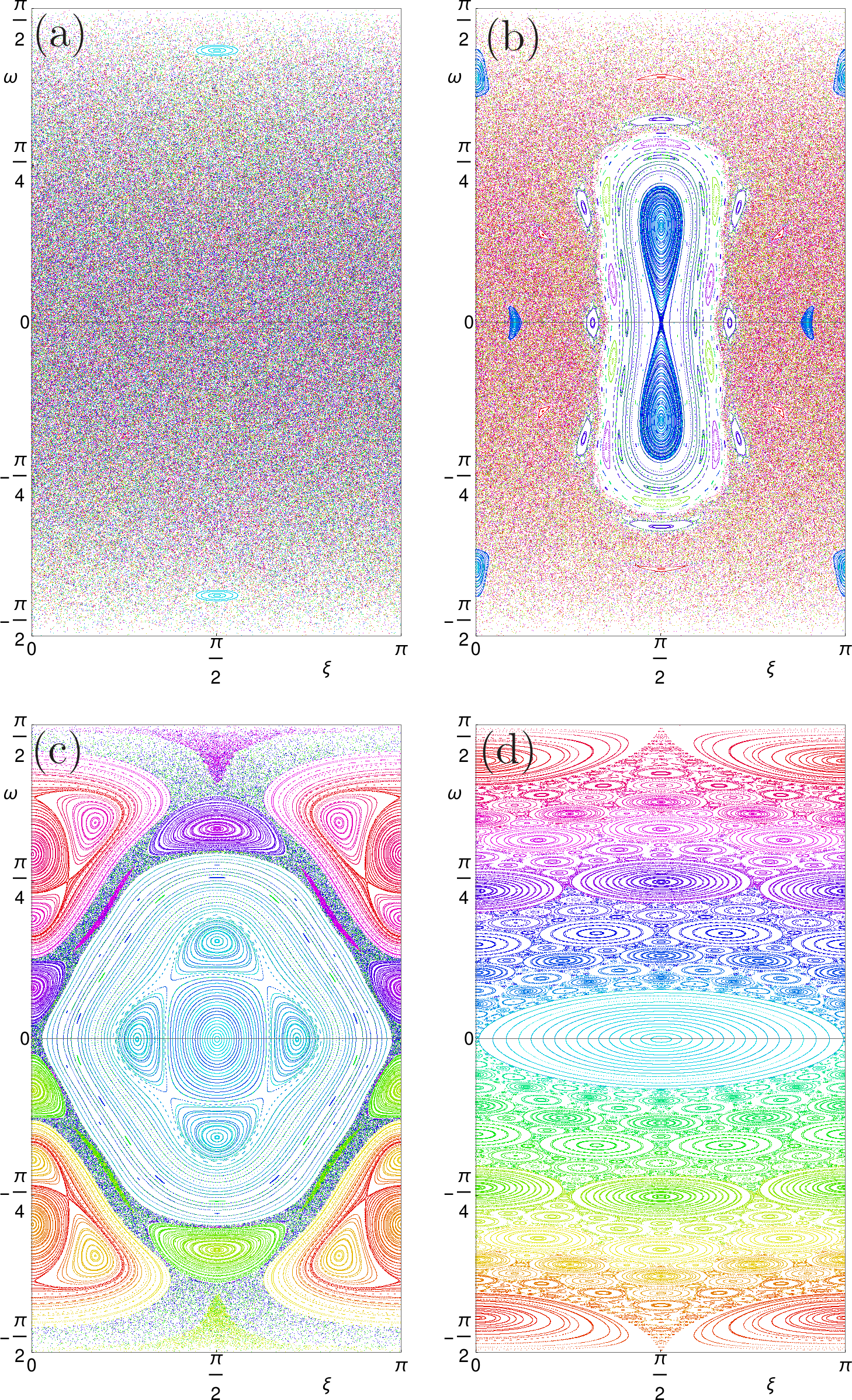}
\caption{The phase space diagram in the plane $(\xi,\omega)$ showing 1000 iterations the Poincar\'e map $M$ for different initial conditions and  $\Phi$ equal to (a) $0.129\,\pi$, (b) $0.461\,\pi$, (c) $0.73\,\pi$ and (d) $0.98\,\pi$. Only one half of the plane $(\xi,\omega)$ is shown; the other half looks the same due to the symmetry of the SWB. The orbits are distinguished by the color hue.}
\label{fig:phasespace1}
\end{figure}

Instead of analyzing the whole trajectory on the SWB, we construct the corresponding Poincar\'e map $M$ by iterating the straight line segments on the disk. This creates a sequence of points $\{(\varth_k,\omega_k);\,k=1,2,\dots\}$ that represent the particle trajectory. 
By a direct calculation we get analytic formulas for the mapping $M$ from one point $(\varth_k,\omega_k)$ to the next one, $(\varth_{k+1},\omega_{k+1})$:
\begin{align} \nonumber
\varth_{k+1}& =  \frac\pi2\sign[\sin(\varth_k-2\omega_k)]\\ & \!\!\!\!\!\!\!\!\!
-\arctan\left\{\frac{\tan\omega_k\sin \Phi}{|\sin(\varth_k - 2\omega_k)|}  - \frac{\cos\Phi}{\tan(\varth_k - 2\omega_k)} \right\}, \label{recur-theta} \\ \nonumber
\omega_{k+1}& = - \arcsin\{\sin \omega_k \cos \Phi \\  & \!\!\!\!\!\!\!\!\!
    +\sign[\sin(\varth_k-2\omega_k)] \cos\omega_k \cos(\varth_k - 2\omega_k) \sin \Phi\}.
\label{recur-omega}\end{align}
It is convenient to introduce an angle $\xi=(\varth-\omega) \,\modd 2\pi$, see Fig.~\ref{fig:manifold}(c)),
where in this paper we define the modulo function ($\modd 2\pi$) such that the result belongs to the interval $(-\pi,\pi]$. It turns out that the variables $(\xi,\omega)$ reflect the symmetry of the problem better than $(\varth,\omega)$, so we will prefer them. Liouville's theorem in the phase space yields an invariant measure $\dd\xi\,\dd(\sin\omega)=\cos\omega\,\dd\xi\,\dd\omega$ of the Poincar\'{e} map $M$. 

\begin{figure}[htb]
\includegraphics[width=84mm]{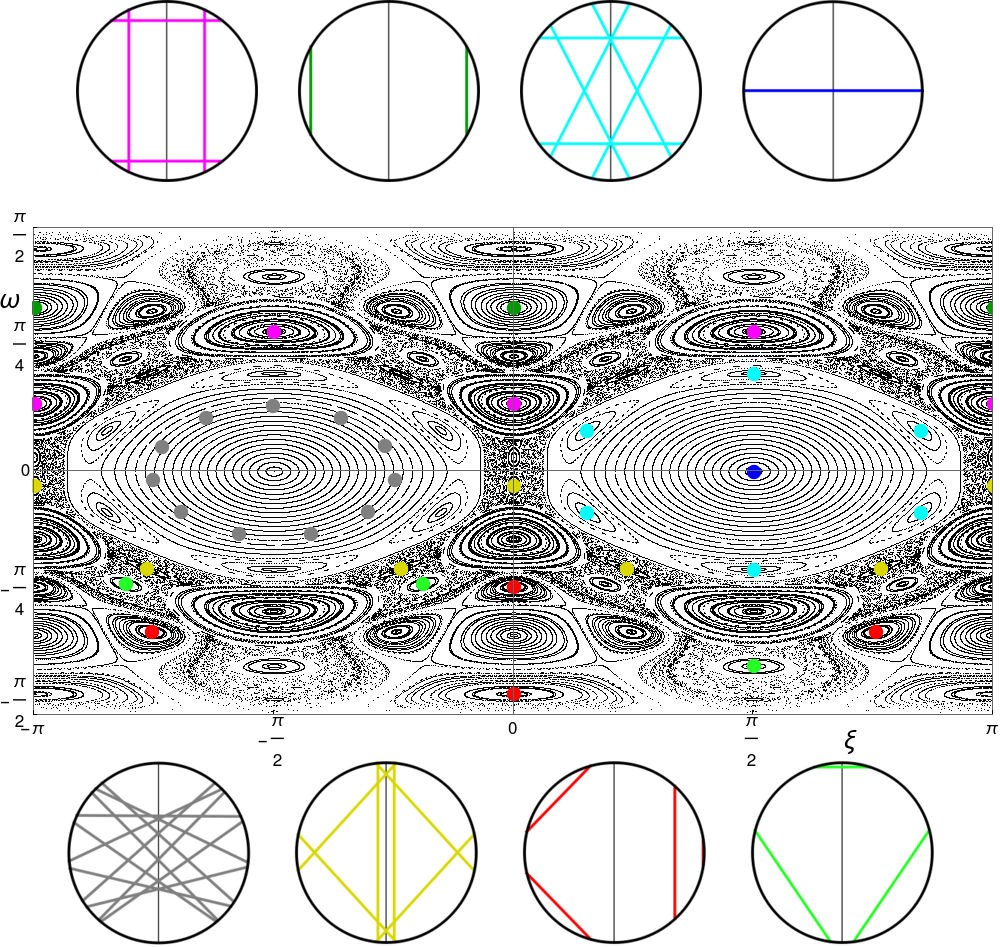}
\caption{The phase space diagram for $\Phi=0.828\,\pi$ with several periodic orbits marked with dots of the same color, and the corresponding trajectory segments on the disk, shown in the same color. The dark-blue dot marks the  point $\Gamma$ discussed in the text.}
\label{fig:map+segments}
\end{figure}

The formulas~(\ref{recur-theta}) and~(\ref{recur-omega}), together with the relations between $(\varth,\omega)$ and $(\xi,\omega)$, i.e., 
\begin{equation}
 \xi=(\varth-\omega) \,\modd 2\pi,\quad \varth=(\xi+\omega) \,\modd 2\pi,
\label{xitheta}\end{equation}
fully describe the behavior of the spherical wedge billiard and conceal enormous complexity whose character varies with the value of the angle $\Phi$. This can be visualized by plotting the phase space diagram in the plane $(\xi,\omega)$, see Fig.~\ref{fig:phasespace1} and supplemental 
\href{https://www.physics.muni.cz/~tomtyc/spherical_wedge_billiard/video1.mp4}{Video 1}. See also supplemental \href{https://www.physics.muni.cz/~tomtyc/spherical_wedge_billiard/video2.mp4}{Video 2} for the correspondence of the disk trajectory segments with the points in the plane $(\xi,\omega)$. 

\begin{figure}[htb]
\includegraphics[width=86mm,angle=0]{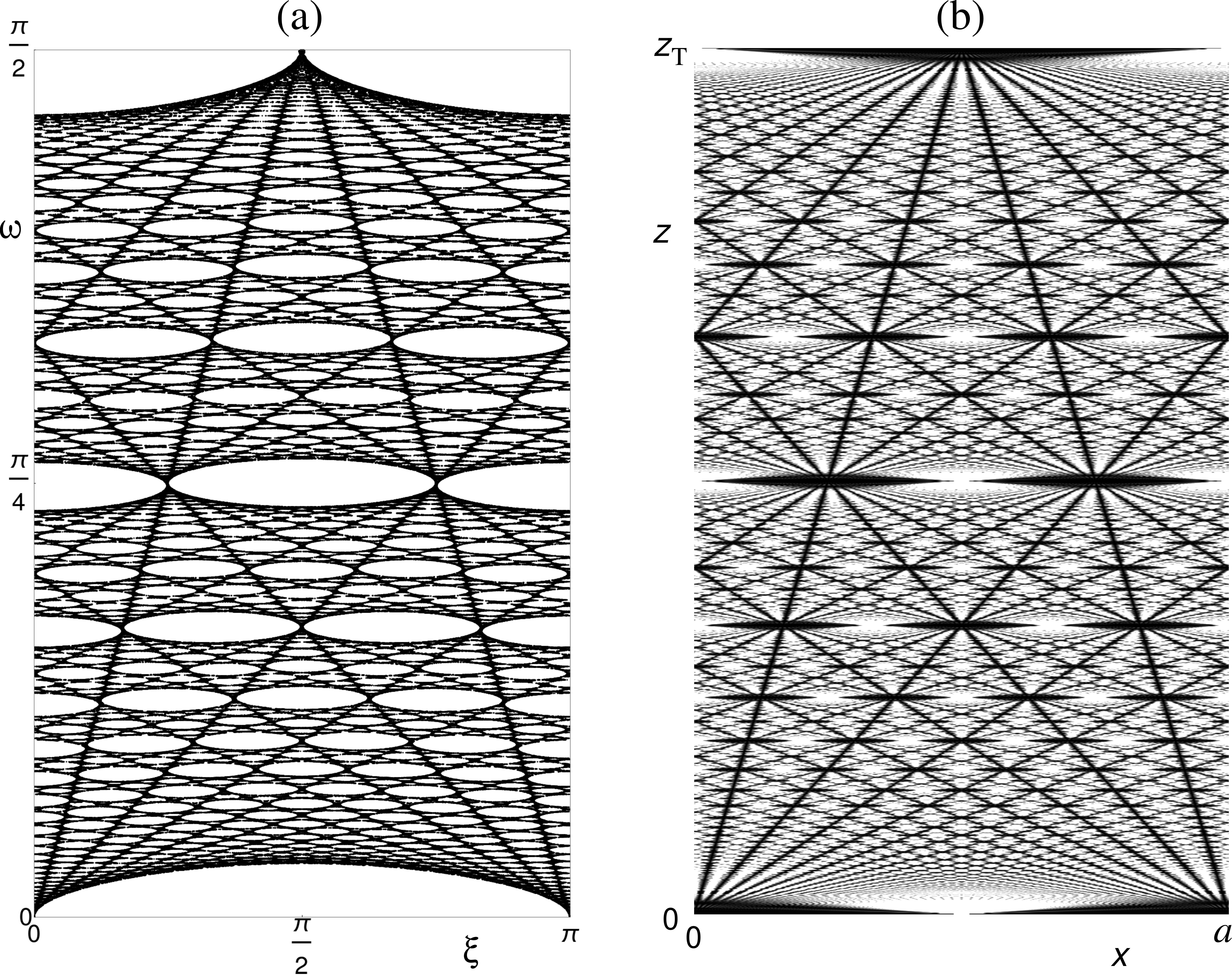}
\caption{Comparison of (a) the phase space diagram for $\Phi=0.997\pi$ (here shown for irregular trajectories starting from the North pole) with (b)
a Talbot carpet~\cite{Berry2001} corresponding to a binary grating with period of $a$ and opening ratio of 1/40. $z_\mathrm T$ is the Talbot distance.}
\label{fig:talbot}
\end{figure}

For small $\Phi$, the behavior is mostly chaotic, with occasional islands of quasiregular motion, see Fig.~\ref{fig:phasespace1}(a-b). For $\Phi\ge\pi/2$, two large regions of quasiregular motion appear, centered at the points $(\varth=\pm\pi/2,\omega=0)$ [one of them can be seen in Fig.~\ref{fig:phasespace1}(c-d)]. As $\Phi$ increases further, more islands of  regular motion appear [see Fig.~\ref{fig:phasespace1}(c-d)] centered around points belonging to periodic orbits, see Fig.~\ref{fig:map+segments}.
When $\Phi$ approaches $\pi$, a very interesting feature can be observed: the plane $(\xi,\eta)$ becomes covered by stacks of ellipses of various sizes that seemingly fill the whole plane, see Fig.~\ref{fig:phasespace1}(d) and Fig.~S1 in Supplemental Material~\cite{supplement}. Still, even then there are tiny regions of irregular behavior between these ellipses, due to fragmentation of the phase space caused by the two poles of the billiard. For $\Phi$ just below $\pi$, the phase space diagram strongly resembles Talbot carpets known from diffraction theory~\cite{Berry2001}, see Fig.~\ref{fig:talbot}. It is out of scope of this Letter to describe all of this complexity; we therefore focus on just a few of the most interesting features.

First, we investigate the situation when $\Phi$ is just slightly smaller than $\pi$, and write it as $\Phi=\pi-\eps$ with $\eps\ll1$. The  the phase space diagram (shown for $\eps=0.02\,\pi$ in Fig.~\ref{fig:phasespace1}(d)) reveals that the plane $(\varth,\omega)$ contains a plethora of stacks of ellipses with similar axes ratios and the same orientations. It is possible to find their approximate equations; the calculation is shown in Supplemental Material~\cite{supplement} for the ellipses centered near the points $(\xi,\omega)=(0,\pi/(2n))$ with $n=2,3,4,\dots$ and it yields the formulas
\begin{align} \label{ellipses1}
 \xi_{2nk} &= A \cos(k\alpha_n+\phi_0),   \\
 \omega_{2nk} &=AR_n \sin(k\alpha_n+\phi_0)+ \frac{\pi}{2n}-\frac\eps{n\sin\frac\pi n\sin\frac\pi{2n}}\,, 
 \label{ellipses2}\end{align}
where $R_n=\sqrt{\eps/\{2n\sin[\pi/(2n)]\}}$ is the ratio of the ellipse axes in the $\omega$ and $\xi$ directions, $\alpha_n=4nR_n$ determines the rate of the point moving along the ellipse, and $\phi_0$ is an initial phase. These equations are valid only up to a certain value of the constant $A$, so a single stack of ellipses fills only a certain region, leaving space for other stacks of ellipses. The higher is the number $n$, the smaller are the stacks. This is the reason why for small $\eps$ the Poincar\'e map resembles Talbot carpets~\cite{Berry2001}, see Fig.~\ref{fig:talbot}: it reflects the structure of rational numbers on the $\omega$ axis in a similar way as is found in the fractional Talbot effect along the light propagation axis, probably including the fractal features; see Supplemental Material~\cite{supplement} for more detail. 

Next we examine the properties of the map near the point  $(\xi,\omega)=(\pi/2,0)$. We will denote this point by $\Gamma$; in Fig.~\ref{fig:map+segments}, it is marked by a dark-blue dot. The path segment on the disk corresponding to $\Gamma$ is a horizontal diameter line going from left to right, and on the spherical part $P_1$ of the spherical wedge billiard, the path follows the equator, $\theta=\pi/2$. Obviously, the point $\Gamma$ is a fixed point of the map $M$. For $\Phi\in[0,\pi/2)$ this point is unstable while for $\Phi\in[\pi/2,\pi)$ it is stable, which can be checked by a direct calculation of the Jacobi matrix eigenvalues. Further, Eqs.~(\ref{recur-theta}), (\ref{recur-omega})  and~(\ref{xitheta}) reveal that for $\Phi\in(0,\pi/2)$, there is a period-2 stable orbit consisting of points $(\xi,\omega)=(\pi/2,\pm\arccos(\tan\frac\Phi2))$, so these points are the fixed points of the mapping $M^2$. As $\Phi$ approaches $\pi/2$, these two points approach each other and merge at the point $\Gamma$ when $\Phi=\pi/2$. Similarly, for $\Phi\in(\pi/2,3\pi/4)$, there is a period-4 stable orbit consisting of points $(\pi/2,\pm\arccos(-\cot\Phi))$ and $(\pi/2\pm\arccos(\cot^2\Phi),0)$, which are hence fixed points of the mapping $M^4$. These points merge together at the point $\Gamma$ when $\Phi=3\pi/4$. It turns out that this is a general behavior, and in the interval $\Phi\in(\pi/2,\pi)$ there are infinitely many such mergers of fixed points of mappings $M^q$ (with different orders $q\in\mathbb N$). We conjecture that an even stronger statement is valid: in any arbitrarily narrow subinterval $I\subset(\pi/2,\pi)$ there are infinitely many such mergers. Conversely, when we decrease the spherical wedge billiard parameter $\Phi$ somewhere within the interval $(\pi/2,\pi)$, the point $\Gamma$ becomes a source of $q$-tuples of stable fixed points of $M^q$ emerging from it, and the set of values of $\Phi$ at which these $q$-tuples emerge is dense. 

Although we have not been able to prove this statement in general, there is a strong analytical and numerical indication that it is valid, which we briefly outline next. To examine the neighborhood of the point $\Gamma$ semianalytically for $\Phi\in(\pi/2,\pi)$, we make substitutions $\xi=\pi/2+\tau$ and $\omega=\eta/\sqrt{-2\cot\Phi}$, and rewrite Eqs.~(\ref{recur-theta} -- \ref{xitheta}) for the variables $(\tau,\eta)$; assuming that they are small, we first keep just terms up to the first order in these variables, which is a good approximation since all the quadratic terms vanish and next order terms are of the third order in $\tau$ and $\eta$. This enables to put the result into the following matrix form,
\begin{equation}
\begin{pmatrix}\tau_{i+1}\\\eta_{i+1}\end{pmatrix}
=\begin{pmatrix}\cos\gamma&-\sin\gamma\\
  \sin\gamma&\cos\gamma\end{pmatrix}\begin{pmatrix}\tau_{i}\\\eta_{i}\end{pmatrix}
\label{matrixeq}\end{equation}
where $\gamma(\Phi)=\arccos(-\cos\Phi-\sin\Phi)\in(0,\pi)$. We see that up to the first order, the iteration in the $(\tau,\eta)$ plane simply corresponds to rotation by the angle $\gamma$ around the origin. Within this approximation, for such values of $\Phi$ that give $\gamma$ as a rational multiple of $\pi$, i.e., $\gamma=\gamma_{pq}\equiv2\pi p/q$ with $p,q\in\mathbb N$, all points $(\tau,\eta)$ would be fixed points of the mapping $M^q$. However, the cubic terms that were neglected in Eq.~(\ref{matrixeq}) slightly alter the transformation, making the rotation angle $\gamma$ slightly decreasing with the radius $r=\sqrt{\tau^2+\eta^2}$, in addition to its dependence on $\Phi$. A closer inspection reveals that in terms of the action of the map $M$ on the rays on the disk, this corresponds simply to spherical aberration. This way, for given integers $p,q$ the stable fixed points of the mapping $M^q$ merge at the point $\Gamma$ for $\Phi=\Phi_{pq}\equiv\gamma^{-1}(\gamma_{pq})$, where $\gamma^{-1}$ denotes the function inverse to the function $\gamma(\Phi)$ defined above; this is illustrated in supplemental \href{https://www.physics.muni.cz/~tomtyc/spherical_wedge_billiard/video3.mp4}{Video 3}. It turns out that the distance of these fixed points from the origin of the plane $(\tau,\eta)$ is almost the same, so they lie very close to a circle centered at the origin, and are almost uniformly spaced along it, thus forming an almost regular $q$-gon. One (in case of odd $q$) or two (for even $q$) vertices of this polygon always lie on the $\eta$ axis. The stable points are interlaced with hyperbolic points, which are also fixed points of the mapping $M^q$; however, these points lie on a slightly smaller circle than the stable points, whereas the relative difference of the radii quickly becomes negligible as $n$ increases. This behavior can be illustrated on the case $p=1, q=4$ described above analytically. 

Moreover, it turns out that the point $\Gamma$ is not the only one having this property of being a source of infinitely many stable fixed points emerging from it as $\Phi$ decreases. In fact, the same holds also for other fixed points of the mapping, in particular the centers of the ellipse stacks discussed above; in this case, however, the positions of these fixed points themselves depend on $\Phi$, as Eqs.~(\ref{ellipses1}) and~(\ref{ellipses2}) show. We thus arrive at a striking behavior of SWB: when changing the parameter $\Phi$, stable fixed points become sources of $q$-tuples of stable fixed points of the ``higher order'', and this way we can continue to infinity. This is a specific form of fractal behavior.

In summary, we have introduced the concept of reflectionless geodesic billiard and presented its remarkable realization, the spherical wedge billiard. Having established an analytic form of the Poincar\'e map that describes it, we have analyzed the system dynamics near a particular fixed point of the map and shown that there is a complex fixed points merging with fractal features, related to spherical aberration. We have also described the system dynamics for the parameter $\Phi$ approaching $\pi$ and shown that the phase space gets covered with stacks of ellipses whose positions reflect the structure of rational numbers; this leads to a striking similarity of SWB phase space diagrams with Talbot carpets that are known to have fractal features. 

The spherical wedge billiard is unique in several aspects: its dynamics can be described analytically but at the same time it is very complex, ranging from chaos to regular behavior and having fractal features. It is related to several rather different fields of physics, namely nonlinear dynamics, optics of geodesic lenses,  diffraction and number theory (both related to Talbot carpets), and quantum physics. Whereas the analysis of the quantum version of SWB is out of the scope of this Letter, it is the obvious next research focus because of the close relation of energy spectrum of a quantized system to trajectories in its classical counterpart. Moreover, the similarity between RGBs and geodesic lenses enables to use the same mathematical tools for describing both systems, e.g. solving the geodesic equations, or the Helmholtz equation, on the curved surface. 

We believe that the reflectionless geodesic billiard may become an important model system for investigating a number of features of dynamical systems, both classical and quantum. Moreover, the recent achievements in experimental investigation of geodesic lenses should soon enable to
explore RGBs experimentally as well.

\end{document}